\documentclass[conference]{IEEEtran}
\IEEEoverridecommandlockouts
\usepackage[utf8]{inputenc}
\usepackage{cite}
\usepackage{amsmath,amssymb,amsfonts}
\usepackage{algorithmic}
\usepackage{graphicx,subfig}
\usepackage{textcomp}
\usepackage{xcolor}

\usepackage{hyperref}
\usepackage{verbatim}
\usepackage{colortbl}

\usepackage{etoolbox}
\makeatletter
\patchcmd{\@verbatim}
  {\verbatim@font}
  {\verbatim@font\fontsize{8}{9}\selectfont}
  {}{}
\makeatother
\makeatletter
\let \@sverbatim \@verbatim
\def \@verbatim {\@sverbatim \verbatimplus}
{\catcode`'=13 \gdef \verbatimplus{\catcode`'=13 \chardef '=13 }} 
\makeatother

\def\BibTeX{{\rm B\kern-.05em{\sc i\kern-.025em b}\kern-.08em
    T\kern-.1667em\lower.7ex\hbox{E}\kern-.125emX}}
\begin{document}

\title{Towards Explaining STEM Document Classification using Mathematical Entity Linking}




\author{\IEEEauthorblockN{1\textsuperscript{st} Philipp Scharpf}
\IEEEauthorblockA{
\textit{University of Konstanz}\\
Konstanz, Germany \\
philipp.scharpf@uni-konstanz.de}

\and
\IEEEauthorblockN{2\textsuperscript{nd} Moritz Schubotz}
\IEEEauthorblockA{
\textit{FIZ Karlsruhe}\\
Karlsruhe, Germany \\
moritz.schubotz@fiz-karlsruhe.de}

\and
\IEEEauthorblockN{3\textsuperscript{rd} Bela Gipp}
\IEEEauthorblockA{
\textit{University of Wuppertal}\\
Wuppertal, Germany \\
gipp@uni-wuppertal.de}

}

\maketitle

\begin{abstract}
Document subject classification is essential for structuring (digital) libraries and allowing readers to search within a specific field.
Currently, the classification is typically made by human domain experts.
Semi-supervised Machine Learning algorithms can support them by exploiting the labeled data to predict subject classes for unclassified new documents.
However, while humans partly do, machines mostly do not explain the reasons for their decisions.
Recently, explainable AI research to address the problem of Machine Learning decisions being a black box has increasingly gained interest.
Explainer models have already been applied to the classification of natural language texts, such as legal or medical documents.
Documents from Science, Technology, Engineering, and Mathematics (STEM) disciplines are more difficult to analyze, since they contain both textual and mathematical formula content.
In this paper, we present first advances towards \emph{STEM document classification explainability} using classical and mathematical Entity Linking.
We examine relationships between textual and mathematical subject classes and entities, mining a collection of documents from the arXiv preprint repository (NTCIR and zbMATH dataset).
The results indicate that mathematical entities
have the potential to provide high explainability as they are a crucial part of a STEM document.

\end{abstract}


\section{Introduction}


Since the earliest known scheme by the Greek Callimachus, a librarian of the Library of Alexandria, there have been many different efforts to classify subject categories in document collections.
Libraries need to be sorted such that readers can search for literature in specified areas or on specific topics of interest. With the rise of digital libraries, machine-readable documents, and Machine Learning methods, human expert classifiers are supported by computers.
However, while humans partly do, machines typically do not explain their classification decisions. Document classification is often a black box and needs more transparency and explainability.
The topic of explainable AI (XAI) has recently gained increasing interest in Machine Learning applications, e.g., in jurisdiction and medicine where the fate of individual humans depends on AI decisions. In addition to global decision tree interpretation, methods to explain the impact of input features on individual predictions, i.e., locally on single data samples, have been developed.

Mathematical documents from Science, Technology, Engineering, and Mathematics (STEM) fields usually contain a large number of mathematical formulae alongside the text. Thus, some of the documents key concepts can be encoded in symbolic, no-textual form. To explain a mathematical document classification, such entities need to be mapped to natural language concept names first. Mathematical Entity Linking (MathEL) is very challenging since the ambiguity of symbols and multitude of equivalent concept representations is even more tremendous than for text.

Until now, there have been only explainable classification approaches for natural language textual documents, which do not contain mathematical (formula) content.
To address this shortcoming, we develop MathEL for STEM document explainability. We examine relations between subject classes and entities statistically and using state-of-the-art classification explainability approaches. We also examine formula identifier symbol-name distributions (relationships between natural and mathematical language in a STEM document) and their entropy.  Identifiers are formula variables with no fixed value\footnote{\url{https://www.w3.org/TR/MathML3/chapter4.html\#contm.ci}}, e.g., some of the quantities (energy, mass, etc.) in physics formulae.

Our key findings are 1) the ambiguity of identifier symbols is distributed across document category classes while their semantic names are more predictable with smaller categorical cross-entropy; 2) an increasing number of unsupervised identifier name augmentation deteriorates the classification accuracy; 3) removing identifier names from the textual document features significantly reduces classification performance; 4) for 80\% of the examined formulae a concept name could be found within a surrounding text window of $\pm 10$ words; and 5) math features are more explainable than text features and a discriminative ranking is more explainable than a frequency ranking in terms of entropy.

The results show that linking mathematical elements (formulae, identifiers) to their concept names and URIs
can enhance the document classification explainability.
Besides, MathIR systems, such as search engines, recommender, and question answering systems, can employ and profit from machine-interpretable labeled formula data, which is generated by MathEL.

\section{Related Work}

The fastly growing and largely available amount of digital documents has led to the development and employment of methods for Automatic Document Classification (ADC)~\cite{DBLP:conf/jcdl/ScharpfSYHMG20}. Documents need to be rapidly categorized, and human domain expert labeling is tedious and time-consuming. ADC approaches are faster and thus less expensive. Furthermore, the algorithms are portable and can be applied to a variety of other applications, such as spam filtering, sentiment analysis, product categorization, speech categorization, author and text genre identification, essay grading, word sense disambiguation, and hierarchical categorization of web pages~\cite{DBLP:journals/eswa/MironczukP18}.

\subsection{Explainable Text Classification}

Although more effective, machine learning classifiers are less interpretable than classical fuzzy logic methods. This led to efforts to develop more explainable classification models.

In 2014, Martens and Provost present methods to address the problem of high corpus dimensionality~\cite{DBLP:journals/misq/MartensP14}. They extend prior models by defining explanation as a minimal set of words or terms without which the predicted class of interest would change. Subsequently, an algorithm to seek such explainer sets is introduced and evaluated in real-world case studies. In the first, web pages are employed for which advertisement decisions depend on the classification. In the second, news-story topic classification is examined and discussed.

Mahoney et al. present a framework for explainable text classification in legal document review~\cite{DBLP:conf/bigdataconf/MahoneyZHGZ19}. They address the problem of the ADC being a `black box' preventing attorneys from understanding why documents are classified as responsive depending on text snippets. The goal is to reduce reviewers' time and increase their classification confidence. Experiments are carried out on a dataset consisting of 688,294 manually labeled by attorneys. The framework could achieve a `rationale recall' of up to 86.68\% for identifying snippets that would explain the classification decision.

Liu et al. develop a new approach that included fine-grained information (textual explanations for the labels) to explicitly generate the human-readable explanations~\cite{DBLP:conf/acl/LiuYW19}. They construct two new datasets containing summaries, rating scores, and fine-grained classification reasons. Subsequently, they conduct experiments on both datasets, claiming that their model outperformed several high-performing neural network baseline systems.

In 2016, Ribeiro et al. present `LIME' - a method for `Local Interpretable Model-agnostic Explanations' to improve human trust in a classifier~\cite{DBLP:conf/kdd/Ribeiro0G16}. They train an interpretable model locally around individual prediction examples to be explained. The task is framed as a submodular optimization problem. They demonstrate the model performance and flexibility on both classical Machine and neural Deep Learning models for text and images.

In 2019, Lundberg et al. present `SHAP' - an approach to improve the interpretability of tree-based models using game-theoretic Shapley values~\cite{DBLP:journals/corr/abs-1905-04610}. They claim to contribute (1) a polynomial-time algorithm to compute optimal explanations, (2) a new type of explanation that directly measures local feature interaction effects, and (3) a new set of tools for understanding global model structure based on combining many local explanations of each prediction. The tools are evaluated to three applied medical Machine Learning problems (classification of risk factors). The open-source SHAP model\footnote{\url{https://github.com/slundberg/shap}} is heavily employed by currently around 2,400 users.


\subsection{Mathematical (STEM) Document Classification}

There are single-label and multi-label, as well as coarse-grained and fine-grained (hierarchical) classification problems. In the case of mathematical documents, the zbMATH library\footnote{\url{https://zbmath.org}} is sorted and labeled using the fine-grained hierarchical `Mathematics Subject Classification (MSC)'\footnote{\url{https://zbmath.org/static/msc2020.pdf}}, which has a long history\footnote{\url{http://www.mi.uni-koeln.de/c/mirror/www.ams.org/msc/msc-changes.html}} with major version publications every ten years (e.g., 2000, 2010, 2020). While the MSC is three-level, the arXiv preprint repository\footnote{\url{https://arxiv.org}} has only a more coarse-grained two-level taxonomy\footnote{\url{https://arxiv.org/category_taxonomy}}.

Classification approaches can be divided into supervised and semi-supervised. The latter type requires to manually label less data. It can be further divided into few-shot learning (FSL), one-shot learning (OSL), and zero-shot learning (ZSL)~\cite{DBLP:journals/corr/abs-1908-09788}, where the number of `shots' signifies the number of examples, for which the model needs to predict new labels. Due to the large number of classes in fine-grained classification schemes, there are often only zero, one, or a few examples available for each class. This motivates the use of ZSL, OSL, and FSL in this scenario.


In 2008, Watt examined the use of relative symbols and expression frequencies to classify a mathematical document according to the MSC scheme~\cite{watt2008mathematical}. It was found that the particular use of symbols and expressions, i.e., their frequency ranking, vary from area to area between different top-level subjects of the MSC 2000. However, the shape of the relative frequency curve was noted to remain the same. It was claimed that the symbol frequency `fingerprints' for the different MSC areas could be used to classify given mathematical documents.

Kuśmierczyk et al. compared hierarchical mathematical document clustering against the hierarchical MSC classification tree~\cite{DBLP:series/sci/KusmierczykLBN13}. They postulated that the hierarchy was highly correlated with the document content. Using publications from the zbMATH database, they aimed at reconstructing the original MSC tree based on document metadata. For the comparison, they developed novel tree similarity measures. The best results were obtained for 3-level hierarchical clustering using bigram encodings.

In 2014, Schoeneberg et al. discussed part-of-speech (POS) Tagging and its applications for mathematics~\cite{DBLP:conf/mkm/SchonebergS14}. Their goal was to adapt NLP methods to the special requirements for STEM document content analysis. They presented a mathematics-aware POS tagger for mathematical publications. It was trained using keyphrase extraction and classification of documents from the zbMATH database. The results showed that while the precision was sufficient (for 26 of the 63 top-level classes higher than 0.75 and only for 4 classes smaller than 0.5), the recall was very low.

In 2017, Suzuki and Fujii presented a structure-based method for Mathematical Document Classification~\cite{DBLP:conf/jcdl/SuzukiF17}. They include structures of mathematical expressions (ME) as classification features combined with the text. They hypothesize that ME would hold important information about mathematical concepts, being a central part of communication STEM fields. Employing 3,339 Q\&A threads from MathOverflow\footnote{\url{https://mathoverflow.net}} and 37,735 papers from the arXiv, they achieve classification F-measures of 0.68 on text and 0.71 on combined text and math encodings.

In 2019, Ginev and Miller performed a supervised Scientific Statement Classification over arXiv.org\cite{DBLP:conf/lrec/GinevM20}. Exploring 50 author-annotated categories, they group 10.5 million annotated paragraphs into 13 classes. Using a BiLSTM encoder-decoder model, a maximum F1-score of 0.91 is achieved. Further, they introduce a lexeme serialization for mathematical formulae and discuss the limitations of both data and task design, complaining the lack of capacity to provide a real human evaluation of the classification results.

In 2020, Scharpf et al. presented large-scale experiments for classification and clustering of arXiv documents, sections, and abstracts comparing encodings of natural and mathematical language~\cite{DBLP:conf/jcdl/ScharpfSYHMG20}. They evaluate the performance and runtimes of selected algorithms, achieving classification accuracies up to 82.8\% and cluster purities up to 69.4\%. Further, they observe a relatively low correlation between text and math similarity, indicating a potential independence of text and formula document features. Moreover, they demonstrate the computer outperforming a human expert in classification performance. Lastly, they discuss the inter-class variance of the math encodings and the need for identifier semantics disambiguation by unsupervised enrichment or supervised annotation.

Most recently, Schbotz et al. presented `AutoMSC' - a system for the automatic assignment of Mathematics Subject Classification (MSC) labels~\cite{DBLP:conf/mkm/SchubotzSTKBG20}. Evaluating the performance of automatic methods in comparison to a human baseline, they find that their best performing method achieving an F1-score of 77.2\%. They claim that using their models, the manual classification effort could be reduced by 86.2\% without losing classification quality.

\subsection{Mathematical Entity Linking}

Entity Linking (EL) is the task of linking textual entities, such as concepts, persons, etc. to unique identifiers (URLs)~\cite{DBLP:journals/ai/HacheyRNHC13}. EL has a variety of Information Retrieval (IR) and Natural Language Processing (NLP) applications, such as semantic search and question answering, text enrichment, relationship extraction, entity summarization, etc.~\cite{DBLP:conf/amw/Rosales-MendezP18}. On the other hand, Mathematical Entity Linking (MathEL) links formulae and their constituting entities (identifiers, operators, etc.) to concept names or Wikimedia URLs~\cite{Scharpf2021}. Among other applications, MathEL allows for formula referencing (math citations) and Mathematical Question Answering (MathQA)~\cite{schubotz2018introducing}.

In 2016, Kristianto et al. propose a learning-based approach to link mathematical expressions to Wikipedia articles by exploiting textual and mathematical document features~\cite{DBLP:conf/icadl/KristiantoTA16,DBLP:conf/wsdm/KristiantoA17}. They introduce a dataset for benchmarking training and testing, achieving a precision of 83.40\%, compared to the 6.22\% baseline performance. One of their findings is that important math expressions can often be found at the top of articles and are highlighted as boxed block-level.

In 2019, `Formula Concepts' (FC) are introduced as collections of equivalent formulae with different notational representations that can be bundled under the same name (and linked to a Wiki resource)~\cite{DBLP:conf/sigir/ScharpfSCG19}. Their goal is to generalize and extend classical textual citation methods to include mathematical formula citations. They call out for a Formula Concept Retrieval challenge with two subtasks: Formula Concept Discovery (FCD) and Formula Concept Recognition (FCR). While the goal of FCD is to analytically or empirically discover a definition of an FC, the purpose of FCR is to match and link math expressions occurring in documents to unique resources (e.g., Wikidata item ID). The authors develop and evaluate first Machine Learning based approaches to tackle the FCD and FCR tasks.

Apart from unsupervised methods for FC retrieval, also supervised approaches were presented. The `AnnoMathTeX' formula and identifier annotation recommender system is designed to disambiguate and match mathematical expressions in Wikipedia articles to Wikidata items~\cite{DBLP:conf/recsys/ScharpfMSBBG19}. The system suggests annotation name and item candidates provided from several sources, such as the arXiv, Wikipedia, Wikidata, or the text that surrounds the formula. A first evaluation showed that in total, 78\% of the identifier name recommendations were accepted by the user.

The study was subsequently extended to an assessment of the community acceptance of the Wikipedia article link and Wikidata item seed edits~\cite{Scharpf2021}. The contributions were accepted in 88\% of the edited Wikipedia articles and 67\% of the Wikidata items. Furthermore, a speedup of the annotation process by a factor of 1.4 for formulae and 2.4 for identifiers was achieved. The recommender system is intended to be soon integrated seamlessly into the Wikimedia user interfaces via a `MathWikiLink' API.


Having reviewed the state of the art in explainable text classification, mathematical (STEM) document classification and mathematical Entity Linking, our goal is to bridge the topics. In this paper, we would like to address the research gap and need for explainable mathematical document classification systems that employ mathematical Entity Linking and assess the relations between document categories and entities.

\section{Methods}

In this paper, we address the information need of document subject category classification interpretability and explainability. Therefore, we analyze the relationships between categories (labels) and entities (features) of a document.


\subsection{Research Questions}

Our research is driven by the following questions:

\begin{enumerate}
    \item How can we quantity and tackle the ambiguity of formula identifier symbols?
    \item Does unsupervised enrichment of identifier names enhance document classification accuracy?
    \item Does unsupervised removal of identifier names from the text deteriorate document classification accuracy?
    \item How well does unsupervised entity linking perform both on text and math elements of a STEM document?
    \item Do text or math document features provide more interpretability or explainability?
\end{enumerate}

Specifically, we would like to know whether it is possible to predict the subject class using formula concept and identifier entities as features. Can formula and identifier entity labels be used to improve classification?
Further, we would like to find out whether there are namespace clusters for identifier annotations per subject class. Is it possible to predict the identifier annotation disambiguation using the subject class and vice versa? Can we employ subject class labels to improve identifier and formula entity link disambiguation?
Moreover, we will explore the quality of inferring relations between arXiv subject classes and MSCs labels.
Lastly, we aim to investigate whether all formula concept entity names appear somewhere in the surrounding text or are hidden in the context, requiring domain knowledge to infer them.

Our long-term goal is to establish a framework to employ entity links as features for document classification and category labels as namespaces for identifier disambiguation.

\subsection{Research Tasks}

To answer our research questions, we set up the following tasks:

\begin{enumerate}
    \item Select documents from arXiv corpus with both arXiv subject class and MSC available (physics domains).
    \item Evaluate the MSC-arXiv category correspondence predictability.
    \item Analyze the class distributions of identifier symbols and names (semantics) and their entropies.
    \item Evaluate the impact of unsupervised semantic identifier enrichment on document classification.
    \item Evaluate the impact of category-concept augmentations on document classification.
    \item Build an Entity Linker (Wikifier) for both natural language (classical EL) and mathematical language (MathEL) entities.
    \item Evaluate different retrieval methods and sources in comparison.
    \item Evaluate the class-entity explainability using different class limits, features, samplings, encodings, classifiers, explainers, and prediction modes.
\end{enumerate}

In our experiments, we vary numerous \textit{evaluation parameters}, such as included subject classes and granularity, ranking size, distribution type, n-gram length, text cleaning, word windows, etc.
We evaluate our experimental results using several different \textit{evaluation metrics}, such as class prediction accuracy, entity prediction relevance and ranking, precision, recall, F1-measure, and distribution entropy.

\section{Evaluation}

\begin{figure}[t]
    \centering
    \includegraphics[width=\columnwidth]{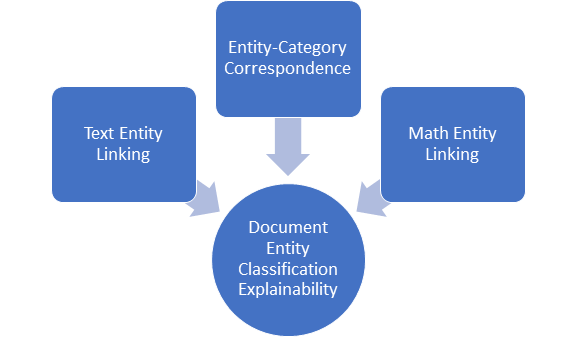}
    \caption{Evaluation workflow. Entity Linking for both textual and mathematical entities and entity-category correspondence is examined as a prerequisite for classification entity explainability.}
    \label{fig:EvalWorkflow}
\end{figure}

In this section, we will present and discuss our evaluation results.
Data, code and result tables are publicly available at \url{https://github.com/AnonymousCSResearcher/STEMdocClassEntityExplainability}.
Figure \ref{fig:EvalWorkflow} shows the workflow of our experiments to examine Entity Linking for both textual and mathematical entities and entity-category correspondence as a prerequisite for classification entity explainability.

We first introduce the NTCIR-11/12 arXiv dataset, which we employ. Subsequently, we examine the correspondence of arXiv and MSC categories. Next, we perform an identifier class semantics distribution analysis. We then test the impact of unsupervised semantic identifier enrichment and category concept augmentations on document classification. Finally, we present Wikification Entity Linking for both natural and mathematical language entities (text and formulae) and explore class-entity correlations to achieve classification explainability.

\subsection{NTCIR arXiv Dataset}

In our experiments, we employ documents (research task 1) from the arXiv NTCIR-11/12 dataset~\cite{DBLP:conf/ntcir/AizawaKOS14}. The dataset is available at \url{http://ntcir-math.nii.ac.jp/data}. It was provided by the National Institute of Informatics Testbeds and Community for Information Access Research Project (NTCIR) and contains 105,120 document section files from scientific articles in English. Each arXiv document in the corpus was divided into paragraphs to make it more digestible for math retrieval tasks. The documents were converted from \LaTeX to an HTML + MathML based TEI\footnote{\url{https://tei-c.org}} format by the KWARC\footnote{\url{http://kwarc.info}} project. The articles were selected from the arXiv categories math, computer science, and physics disciplines. The documents contain about 60 million mathematical formulae, including monomial expressions, e.g., $x$ or $t^2$. The disc size of the dataset is about 174 GB uncompressed and is intended to be used for Information Retrieval research tasks, such as Natural Language Processing, text analysis, and mathematical expression tree structure search\footnote{\url{http://ntcir-math.nii.ac.jp/ntcir11-math/task-overview}}.

\subsection{MSC-arXiv Category Correspondence}

In our first experiment, we analyze the correspondence of arXiv and MSC subject class categories (research task 2) since the two different schemas need to be consolidated. Document classification explainability can refer to either one of those schemas (and others if available). We approach this task two-fold: 1) inferring relations from a constructed co-occurrence matrix, 2) using TF-IDF encodings and a Logistic Regression classifier. For 258,830 documents, both arXiv category and MSC classification labels were available, which could so be matched per document. We created a correspondence table with arXiv ID, zbMATH ID, MSC categories, and arXiv categories. The first two columns contain identification numbers to locate the documents in the arXiv and zbMATH databases, respectively. The last two columns contain the classification categories in the arXiv and MSC schemes, respectively. The MSC number has 5 digits (three-level taxonomy, e.g., `85-A-05' for `galactic and stellar dynamics astrophysics') while the arXiv subject contains subject and subcategory (two-level taxonomy, e.g., `astro-ph.SR' for `solar and stellar astrophysics').

From the co-occurrence matrix, we derive two uncertainty measures. For each distribution, we first calculate the \textit{(Shannon) entropy uncertainty} as
\begin{align*}
S = - \sum_i p_i \cdot \log_2 \left(p_i\right)
\end{align*}
Secondly, we calculate the \textit{margin uncertainty} as the probability difference between the first and second highest class count. With increasing margin, the (mixup) uncertainty decreases and classification accuracy increases. For the arXiv subject class distributions the results are 1) entropies: (mean, max) = (1.86, 3.11), and 2) margins: (mean, max) = (1.00, 1.00). For the MSC category distributions the results are 1) entropies: (mean, max) = (1.67, 3.59), and 2) margins: (mean, max) = (0.02, 0.99). It is evident that from the co-occurrence frequency statistics, arXiv categories can be predicted more accurately than MSCs, as the significantly larger mean margin indicates.

Next, we investigate whether this result is qualitatively reproduced by a trained classifier. We tackle the multi-label data as multiple repeated single-label instances. This means that if for one document there are, e.g., two-class labels to predict, we add them as two separate data points.
We employ a Logistic Regression classifier on TF-IDF encodings to compare different prediction directions (from MSC to arXiv and vice versa), label modes (single-label vs. multi-label), and levels of granularity (coarse, fine). Table \ref{tab:arXivMSCpred}) shows the results of this experiment. The result from the first experiment that the prediction of arXiv categories from MSCs is more accurate is qualitatively reproduced. It can be explained by the fact that at the deepest taxonomic level, the total number of MSCs (6202) is much larger than the number of arXiv categories (156). Due to this large number of MSCs, we focus on arXiv categories in subsequent explainability experiments, since for the latter, we have more data for each class.

\begin{table}[]
\caption{Prediction accuracy of arXiv subject classes from MSC labels and vice versa using a Logistic Regression classifier on TF-IDF encodings, while comparing different label modes and levels of granularity.}
\centering
\begin{tabular}{|l|l|l|}
\hline
\textbf{Prediction/granularity} & \textbf{Coarse} & \textbf{Fine} \\ \hline
single-label                    & \multicolumn{2}{l|}{}           \\ \hline
arXiv from MSC                          & 0.87            & 0.69          \\ \hline
MSC from arXiv                            & 0.61            & 0.31          \\ \hline
multi-label                     & \multicolumn{2}{l|}{}           \\ \hline
arXiv from MSC                           & 0.72            & 0.44          \\ \hline
MSC from arXiv                             & 0.52            & 0.42          \\ \hline
\end{tabular}
\label{tab:arXivMSCpred}
\end{table}

Comparing the classifier predictions to Argmax predictions (most frequent class is predicted) from the co-occurrence matrix, we find only 3 matches and 153 mismatches for the 156 arXiv categories. This can potentially be explained by the difference that cross-correlations are taken into account by the encodings and the classifier but not included in the occurrence frequency statistics.
One could also perform multi-label classification with different weights for the first, second, third, etc., most probable prediction. However, it is a priori unclear how many labels are necessary for each individual document (it can vary), and even a confidence threshold would not reasonably mitigate the problem since it is difficult or impossible to determine.

\subsection{Identifier Class Semantics Distributions}

A mathematical formula typically consists of `identifiers', such as $E$, $m$, and $c$ for $E=mc^2$. In addition to the formula concept, these identifiers are also linkable mathematical entities. Unfortunately, the symbols are highly ambiguous. For example, the variable $x$ is used in a variety of different contexts besides the most common meanings of `unknown', `coordinate', and `vector'.
For document classification explainability, it is essential to know how the meanings (in the following named `semantics') are distributed over subject classes or categories. To examine this, we address research question 1 by research task 3.

\begin{figure*}[h]
    \centering
    \subfloat{\includegraphics[width=0.49\textwidth]{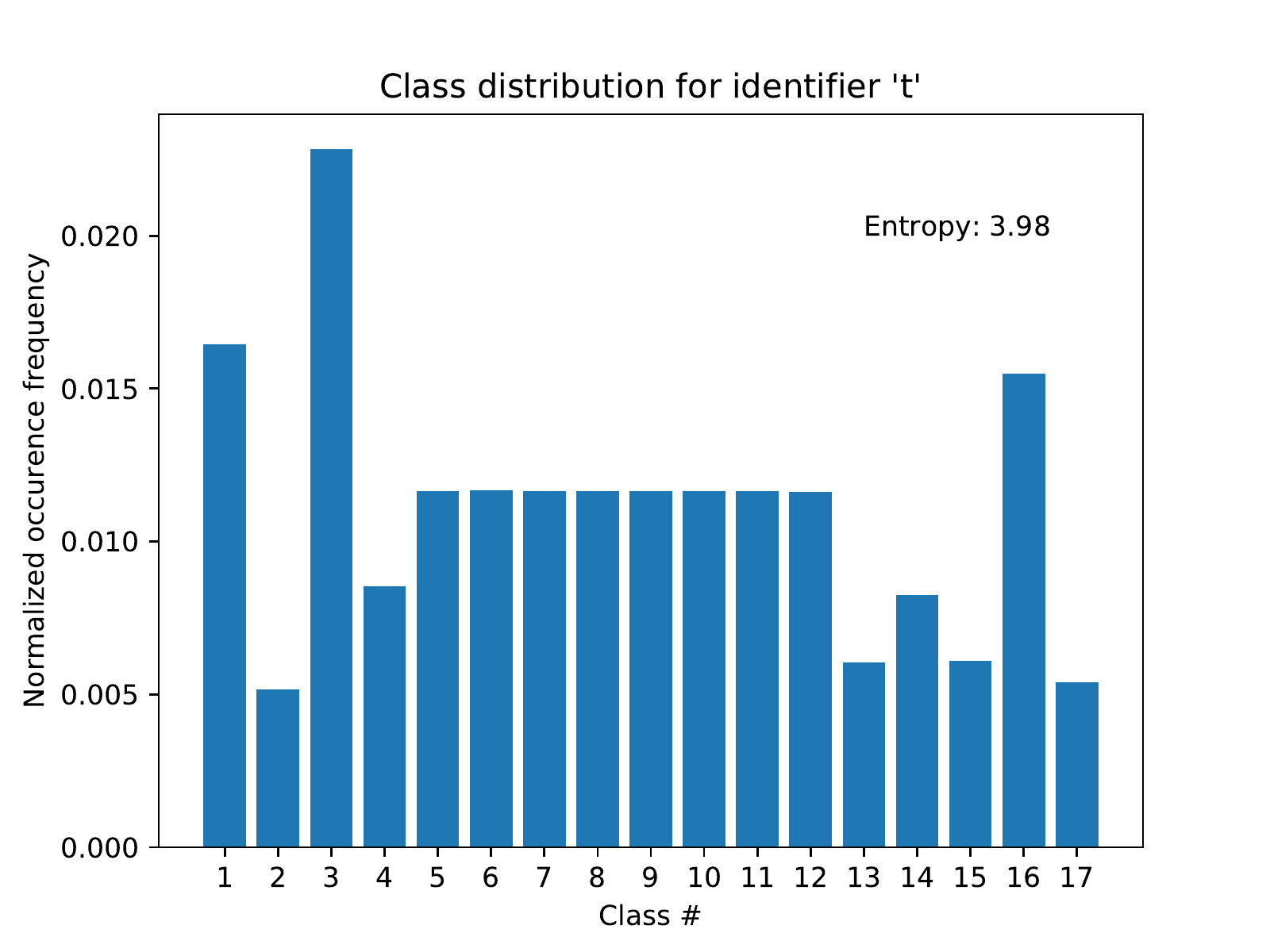}}
    \subfloat{\includegraphics[width=0.49\textwidth]{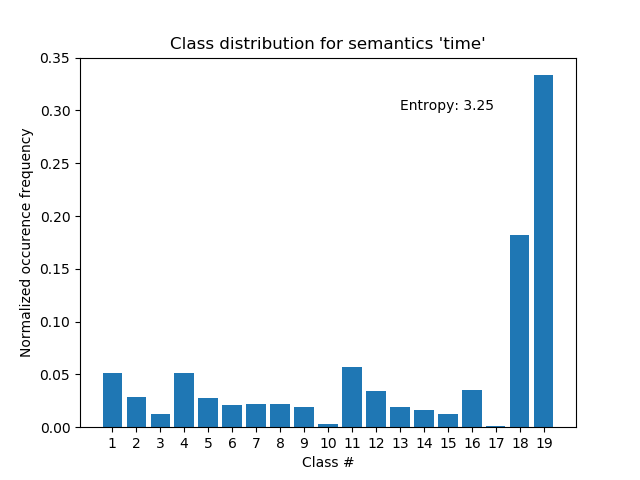}}
    \caption{Class distribution for identifier symbol `t' in comparison to semantics `time', which has a lower information entropy.}
    \label{fig:Class_distribution_t_time}
\end{figure*}

Figure \ref{fig:Class_distribution_t_time} shows a comparison of the class distribution of the identifier symbol $t$ (left) to the semantic name `time' (right), which was found to be the most used in physics on the employed NTCIR test dataset. The uniform normalized occurrence frequency indicates that for the symbol (left), there are more equal values for several classes. This means that the use of the symbol is not a good predictor of the class to provide explainability. On the other hand, for the semantic name (right), there are sharper peaks (at the right end). Calculating the Shannon information distribution entropies, we can support this finding by a smaller value (3.25 \textless ~ 3.98).

\begin{table}[h]
    \caption{Comparison of identifier symbol (e.g., `m') and name (e.g., `mass') distribution entropies.}
    \resizebox{0.48\textwidth}{!}{
    \begin{tabular}{|c|c|c|c|}
         \hline
         \textbf{Feature/measure} & Min (entropies) & Mean (entropies) & Max (entropies) \\
         \hline
         Identifier symbols & 2.01 & 3.77 & 4.07 \\
         \hline
         Identifer names & 0.00 & 0.56 & 4.24 \\
         \hline
    \end{tabular}}
\label{tab:identifier_symbol-name_distribution-entropies}
\end{table}

To generalize this example observation, we calculate the entropies for the whole corpus (Table \ref{tab:identifier_symbol-name_distribution-entropies}). The mean of the entropies of the identifier symbols is almost 7 times larger than the mean of the entropies of the identifier names.
This means that the identifier semantics can more 'sharply' be determined by or explain the subject class, whereas identifier symbols widely spread over all subject classes.
This is not surprising considering that word sets are much larger than identifier symbol sets (Latin and Greek alphabet), reflecting symbol ambiguity and the need for semantic enrichment, MathEL, and an identifier semantics vocabulary.

\begin{table*}[h]
\caption{Scheme of the identifier semantics class statistics library. For each distribution, the number of documents containing the respective combination - e.g., identifier-class - is displayed.}
\centering
\begin{tabular}{|l|l|}
\hline
\textbf{Order Scheme}                   & \textbf{Example}                                                                   \\ \hline
Class distribution                      & \{'quant-ph': 6082, 'math': 3997, 'physics': 438, ...\}                            \\ \hline
Class-identifier distribution           & \{'math': \{'e': 61, 'T': 45, 'M': 35, ...\}, ...\}                                \\ \hline
Class-semantics distribution            & \{'physics': \{'field': 50, 'function': 30, 'constant': 25, ...\}, ...\}           \\ \hline
Identifier-class distribution           & \{'t': \{'quant-ph': 100, 'math': 33, 'physics': 10, ...\}, ...\}                  \\ \hline
Identifier-semantics distribution       & \{'t': \{'time': 831, 'coordinate': 215, 'function': 182, ...\}, ...\}             \\ \hline
Identifier-class-semantics distribution & \{'t': \{'quant-ph': \{'time': 55, 'system': 9, 'function': 9, ...\}, ...\}, ...\} \\ \hline
Semantics-class distribution            & \{'time': \{'quant-ph': 314, 'nlin': 30, 'physics': 25, ...\}, ...\}               \\ \hline
Semantics-identifier distribution       & \{'time': \{'t': 95, 'tau': 61, 'T': 32, ...\}, ...\}                              \\ \hline
Semantics-class-identifier distribution & \{'time': \{'quant-ph': \{'t': 55, 'tau': 26, 'T': 24, ...\}, ...\}, ...\}         \\ \hline
\end{tabular}
\label{tab:Identifier-semantics-class statistics library}
\end{table*}

Table \ref{tab:Identifier-semantics-class statistics library} summarizes our identifier statistics library catalog containing attributions between classes, identifier symbols (`identifier'), and names (`semantics').
For example, the `semantics-identifier distribution' shows that the semantics `time' denotes 95 times the identifier symbol `t', 61 times `tau', etc. Conversely, the identifier symbol `t' is denoted 831 times with the semantics `time', 215 times with `coordinate', etc.

\subsection{Unsupervised Semantic Identifier Enrichment for Document Classification}

As a prerequisite for evaluating the classification impact of formula MathEL, we assess the effect of identifier MathEL (research task 4) first.
In this subsection, we perform experiments to answer research questions 2 and 3. First, we would like to know whether the unsupervised (since supervised is costly and not scalable) enrichment of identifier names improves or deteriorates document classification accuracy.
After parsing the documents from the NTCIR dataset, identifier symbols are extracted from their formula MathML markup. Subsequently, the (Latin and Greek letters) are mapped to their textual meaning, e.g., 'energy' for the $E$ in $E=mc^2$. For the identifier augmentation, we employ the following sources:

\begin{itemize}
    \item \textbf{Wikipedia}: identifier symbol-value relations retrieved from surrounding text statistics of Wikipedia articles\footnote{\url{https://en.wikipedia.org/w/index.php?title=User:Physikerwelt&oldid=738857609}};
    \item \textbf{arXiv}: identifier symbol-value relations retrieved from occurrence frequency statistics in arXiv documents;
    \item \textbf{Wikidata}: identifier symbol-value relations retrieved from mathematical Wikidata items via SPARQL queries\footnote{\url{https://query.wikidata.org/}}.
\end{itemize}

We employ the following subject class selection: `cs' (computer science), `physics', `astro-ph', `cond-mat', `gr-qc', `hep-lat', `hep-ph', `hep-th', `quant-ph' (physics), `math-ph', `math', `nlin', `alg-geom', `q-alg' (mathematics).

Using a Logistic Regression classifier on TF-IDF encodings, we find out that the 80.6 accuracy of text-only is drastically reduced to 23.0 when including identifier symbols. However, identifier name augmentations almost double the latter result to accuracy values around 0.5, as shown in Table \ref{tab:identifier_semantification}.

\begin{table}[h]
\caption{Classification accuracy using Logistic Regression on TF-IDF encodings of NTCIR arXiv document texts that were extended by unsupervised identifier semantification names provided by the AnnoMathTeX API \cite{DBLP:conf/recsys/ScharpfMSBBG19,Scharpf2021}.}
\centering
\begin{tabular}{|l|l|l|}
\hline
\textbf{Source/Cutoff} & \textbf{Top 3} & \textbf{Top 5} \\ \hline
arXiv                  & 0.53           & 0.50           \\ \hline
Wikipedia              & 0.49           & 0.46           \\ \hline
Wikidata               & 0.51           & 0.49           \\ \hline
\end{tabular}
\label{tab:identifier_semantification}
\end{table}

We compare enriching with top 3 to top 5 candidates and conclude from the observation that the former slightly outperforms the latter that it is better to have less but more meaningful augmentations. We also note that the arXiv slightly outperforms the other sources. However, the differences between all measured values are very small.
We conclude that MathEL for identifiers maybe not necessary for explainability as it deteriorates the classification accuracy.

\subsection{Category-Concept Augmentations}

Having investigated the effect of identifier name augmentations (research task 4), we continue exploring the impact of formula concept name augmentations (research task 5). Specifically, we employ labelled concept-(sub)category mappings provided by Wikipedia authors\footnote{\url{https://en.wikipedia.org/w/index.php?title=Outline\_of\_physics&oldid=997028605\#General_concepts_of_physics}}.

As there is no cross-discipline data available, in this experiment we employ a smaller subject class selection only from physics: `astro-ph', `cond-mat', `gr-qc', `hep-lat', `hep-ph', `hep-th', `quant-ph'.

\begin{table}[h]
\caption{Classification accuracy of Logistic Regression on TF-IDF encodings of different document content types (text, math) and combinations thereof (plus, minus).}
\centering
\begin{tabular}{|l|l|l|l|l|}
\hline
\textbf{Encoded} & \textbf{Text} & \textbf{Math} & \textbf{Text + Math} & \textbf{Text - Math} \\ \hline
Accuracy         & 0.73          & 0.60          & 0.60                 & 0.19                 \\ \hline
Duration         & 1.0           & 0.21          & 0.20                 & 0.04                 \\ \hline
\end{tabular}
\label{tab:concept-category_augmentations}
\end{table}

Adding the concept-category augmentations does deteriorate the full text classification accuracy of 73.3 to 60.0 of employing the concept-category augmentations as features alone (see Table \ref{tab:concept-category_augmentations}). Interestingly, removing the augmentations from the full text reduces the classification performance to 19.0 only. This means that the augmentations seem to have some explanatory value.
Further investigations show that all of the augmented categories are present in at least one of the documents per class.

\subsection{Wikisource Entity Linking (Wikification)}

The next step in our experiments to examine classification explainability by entity linking is to evaluate natural and mathematical Wikisource entity linker for STEM document Wikification.
We, therefore, investigate research question 4 by carrying out research tasks 6 and 7.
From the NTCIR dataset, we select a random document with both arXiv category and MSC available.

\paragraph{Natural Language Entities}

First, we evaluate different entity linking methods or sources, such as a Wikipedia article name dump\footnote{\url{http://dumps.wikimedia.org/enwiki/latest/enwiki-latest-all-titles-in-ns0.gz}}, and Wikidata Pywikibot\footnote{\url{https://github.com/wikimedia/pywikibot}} or SPARQL retrieval. The unsupervised approaches predict Wikipedia article names and URLs or Wikidata item names and QIDs.

\begin{table*}[ht]
\caption{Excerpt (7 out of 198 tuple lines) from the evaluation of Natural Language Entity Linking to Wikipedia and Wikidata for concepts in a selected abstract from the NTCIR arXiv dataset. A binary classification is made for Wikidump name (eval1) and URL (eval2), Pywikibot retrieved item (eval3) and QID (eval4), and SPARQL retrieved item (eval5) and QID (eval6). The respective values for precision, recall, and F1 measure are displayed in bottom lines (unlemmatized / lemmatized).}
\resizebox{\textwidth}{!}{
\begin{tabular}{|l|l|l|l|l|l|l|l|l|l|}
\hline
\textbf{Tuple}      & \textbf{Relevance} & \textbf{Wikipedia URL}       & \textbf{Wikidata QID} & \textbf{eval1} & \textbf{eval2} & \textbf{eval3} & \textbf{eval4} & \textbf{eval5} & \textbf{eval6} \\ \hline
find that           & 0                  & -                            & -                     & FP             & FP             & FP             & TN             & TN             & TN             \\ \hline
that the            & 0                  & -                            & -                     & TN             & TN             & TN             & TN             & TN             & TN             \\ \hline
the required        & 0                  & -                            & -                     & TN             & TN             & TN             & TN             & TN             & TN             \\ \hline
required velocity   & 1/2                & ...wiki/Velocity             & Q11465                & FN             & FN             & FN             & FN             & FN             & FN             \\ \hline
velocity dispersion & 1                  & ...wiki/Velocity\_dispersion & Q637450               & TP             & TP             & TP             & TP             & TP             & TP             \\ \hline
dispersion is       & 1                  & ...wiki/Dispersion\_relation & Q590051               & FN             & FN             & FN             & FN             & FN             & FN             \\ \hline
is of               & 0                  & -                            & -                     & TN             & TN             & TN             & TN             & TN             & TN             \\ \hline
of the              & 0                  & -                            & -                     & TN             & TN             & TN             & TN             & TN             & TN             \\ \hline \hline
\multicolumn{1}{l}{l} & \multicolumn{1}{l}{l} & \multicolumn{1}{l}{l} & \multicolumn{1}{|l|}{\textbf{Precision}} & 0.63 / 0.67
           & 0.70 / 0.67
            & 1.00 / 0.44
            & 1.00 / 0.44
            & 1.00 / 0.30
            & 1.00 / 0.30
            \\ \cline{4-10}
\multicolumn{1}{l}{l} & \multicolumn{1}{l}{l} & \multicolumn{1}{l}{l} & \multicolumn{1}{|l|}{\textbf{Recall}} & 0.15 / 0.08
           & 0.21 / 0.08
           & 0.01 / 0.56
           & 0.19 / 0.56
           & 0.04 / 0.26
           & 0.21 / 0.26
           \\ \cline{4-10}
\multicolumn{1}{l}{l} & \multicolumn{1}{l}{l} & \multicolumn{1}{l}{l} & \multicolumn{1}{|l|}{\textbf{F1-measure}} & 0.24 / 0.14
           & 0.32 / 0.14
           & 0.02 / 0.49
           & 0.32 / 0.49
           & 0.08 / 0.28
           & 0.35 / 0.28
           \\ \cline{4-10}
\end{tabular}}
\label{tab:nat.el}
\end{table*}

For each tuple in the abstract of the selected document, its relevance (0 or 1, 1/2 if only one of the two words was relevant) as a candidate entity to be linked was manually assessed by a domain expert. Table \ref{tab:nat.el} shows some example lines from our 
evaluation of the EL predictions. The binary classification (true positive: TP, false positive: FP, false negative: FN, true negative: TN) is shown for Wikidump name (eval1) and URL (eval2), Pywikibot retrieved item (eval3), and QID (eval4), and SPARQL retrieved item (eval5) and QID (eval6). The comparison is made with the manual relevance assessment to evaluate the quality of the EL predictions. It is classified FN if the content (page, item, URL, QID) is wrong. While FP and TN can be assessed automatically, TP and FN need to be controlled using human supervision. The links can potentially be used as Wikipedia article enrichment candidates. At the bottom, precision, recall, and F1-measures are displayed for each eval mode (with a distinction between unlemmatized / lemmatized), calculated as

\begin{align*}
  \text{precision} &= \frac{TP}{TP + FP}, \\
     \text{recall} &= \frac{TP}{TP + FN}, \ \text{and} \\
     F1 &= 2 \cdot \frac{\mathrm{precision} \cdot \mathrm{recall}}{ \mathrm{precision} + \mathrm{recall}}.
\end{align*}

Stopwords like `the', `of', `is', `are', etc. are not counted in the classification of relevance (especially if 1/2). Stopword removal and lemmatization (e.g., mapping `vortices' to `vortex') is made using the nltk\footnote{\url{https://www.nltk.org}} python library. We introduce stopword removal and lemmatization because without the precision is high and the recall is low. The number of FP being low indicates that not many non physics-related entities were retrieved, while a high number of FN means that many entities were missed out because of conjugation.

Table \ref{tab:nat.el} reveals that unlemmatized, the SPARQL retrieval from Wikidata (eval6) performs best (highest mean among measures) and Wikidump retrieval (eval1) worst. For the lemmatized case, we find that the mean precision and recall are very similar. The pywikibot item name and QID retrieval (eval 3 and 4) perform best. At the transition from unlemmatized to lemmatized, the precision decreases (mean: 0.47 \textless ~ 0.88) but recall increases (0.30 \textgreater ~ 0.14). The lemmatized mode has a better F1 than unlemmatized in both mean (0.30 \textgreater ~ 0.22) and max (0.49 \textgreater ~ 0.35). The full lists of results can be found in the repository.






\paragraph{Mathematical Language Entities}

Having evaluated natural language entity linking, we now examine the feasibility of mathematical Entity Linking on the selected STEM document from the NTCIR arXiv dataset. Here, the evaluation procedure using manual examinations by our domain expert is three-fold. We first identify relevant formula concept entities as linking candidates in the introduction of the analyzed physics paper. Second, we name them and retrieve the corresponding Wikidata item and QID linking (if available). Third, we spot relevant name candidates in a window of 10 words in the surrounding text before and after the given formula.
We define a score as 0 if `irrelevant', 1 if `partly relevant', and 2 if `highly relevant'.

\begin{table*}[ht]
\caption{Excerpt from the evaluation of mathematical Entity Linking to Wikipedia and Wikidata for Formula Concepts in a selected article from the NTCIR arXiv dataset. N-grams are taken up to N=3, and a score is given as 0 for `irrelevant', 1 for `partly relevant', and 2 for `highly relevant'. The rank is positive if the concept name appears before the formula and negative if after in the text word window.}
\resizebox{\textwidth}{!}{
\begin{tabular}{|l|l|l|l|l|l|}
\hline
Formula Concept ID & Formula Concept Name                   & N-gram & (Score, Rank) & Wikipedia article                              & Wikidata item \\ \hline
1                  & Gross-Pitaevski equation               & 3      & (2,-8)        & Gross–Pitaevskii\_equation                     & Q910667       \\ \hline
2                  & Circulation in condensate              & 3      & (0,-)             & -                                              & -             \\ \hline
2                  & Circulation                            & 1      & (1,-6)        & Circulation\_(physics)                         & Q205880       \\ \hline
2                  & Condensate                             & 1      & (1,-4)        & Bose–Einstein\_condensate                      & Q46202        \\ \hline
3                  & Axion mass                             & 2      & (1,-7)        & Axion                                          & Q792548       \\ \hline
3                  & Axion                                  & 1      & (1,-7)        & Axion                                          & Q792548       \\ \hline
3                  & Mass                                   & 1      & (1,-6)        & Mass                                           & Q11423        \\ \hline
4                  & Gross-Pitaevski equation               & 3      & (2,-6)        & Gross–Pitaevskii\_equation                     & Q910667       \\ \hline
5                  & General solution                       & 2      & (2,-9)        & Gross–Pitaevskii\_equation                     & Q910667       \\ \hline
6                  & Spherical Bessel function              & 3      & (2,-9)        & Bessel\_function\#Spherical\_Bessel\_functions & -             \\ \hline
6                  & Spherical                              & 1      & (1,-9)        & Sphere                                         & Q12507        \\ \hline
6                  & Bessel function                        & 2      & (1,-8)        & Bessel\_function                               & Q219637       \\ \hline
6                  & Bessel                                 & 1      & (1,-8)        & Friedrich\_Bessel                              & Q75845        \\ \hline
6                  & Function                               & 1      & (1,-7)        & Function\_(mathematics)                        & Q11348        \\ \hline
7                  & Energy density of the axion condensate & 4      & (1,-10)       & Energy\_density                                & Q828402       \\ \hline
7                  & Energy density                         & 2      & (0,-)             & Energy\_density                                & Q828402       \\ \hline
7                  & Energy                                 & 1      & (0,-)             & Energy\#Cosmology                              & Q11379        \\ \hline
7                  & Density                                & 1      & (0,-)             & Density                                        & Q29539        \\ \hline
7                  & Axion                                  & 1      & (1,-10)       & Axion                                          & Q792548       \\ \hline
7                  & Condensate                             & 1      & (1,-9)        & Bose–Einstein\_condensate                      & Q46202        \\ \hline
\end{tabular}}
\label{tab:math.el}
\end{table*}

Table \ref{tab:math.el} contains an excerpt of the results of our assessment. It shows 20 of the 56 spotted entity link candidates (for 23 identified formula concepts). For each, the availability of a corresponding Wikipedia article and Wikidata item was checked. We found 51/56 = 91\% for the items and 53/56 = 95\% for the articles. Further, in 43/56 = 77\% of the cases, the surrounding text contained name candidates that can be selected by the entity linker and connected to Wikipedia and/or Wikidata. Among them, 12 were `highly relevant'. For 13 formula concepts, the name was `hidden' in the context, requiring domain knowledge to infer them, which is impossible to do automatically by an entity linker.

\subsection{Class-Entity Explainability}

Finally, we will examine the potential for class-entity explainability. We investigate whether text or math features provide more interpretability (research question 5) using different class limits, features, samplings, encodings, classifiers, explainers, and prediction modes (research task 8).






In our analysis, we employ the following experimental parameter variations:

\begin{enumerate}
    \item Batch: Class limit.
    \item Features: Text, math.
    \item Sampling: Most frequent, most discriminative.
    \item Encoding: TF-IDF, Doc2Vec.
    \item Classifier: LogReg, SVC.
    \item Explainer: LIME, SHAP.
    \item Prediction: Classes from entities (ClsEnt), entities from classes (EntCls).
\end{enumerate}

We achieve the best results using a LIME explainer applied to a LogReg classifier on TF-IDF encodings (default scikit-learn\footnote{\url{https://scikit-learn.org}} configurations).
To assess the potential of entity-category explainability, we examine the following four distributions:

\begin{enumerate}
    \item `MFreqText': Most frequent entities in the textual content.
    \item `MFreqMath': Most frequent entities in the mathematical content.
    \item `MDiscText': Most discriminative entities in the textual content.
    \item `MDiscMath': Most discriminative entities in the mathematical content.
\end{enumerate}

From the NTCIR corpus, we select the following classes: `astro-ph', `cond-mat', `gr-qc', `hep-lat', `hep-ph', `hep-th', `math-ph', `nlin', `quant-ph', `physics'].
%
We then calculate entropies for class entity distributions to find out which setting has the most predictive power and can be employed for document explainability.

\begin{table}[]
\caption{Entropies for class and entity distributions using documents from 10 different arXiv subject classes. The most frequent (MFreq) and most discriminative (MDisc) predictions of classes from entities (ClsEnt) or vice versa (EntCls) are evaluated for textual (Text) and mathematical (Math) entities.}
\centering
\begin{tabular}{|l|l|}
\hline
\textbf{Mode}   & \textbf{Entropy} \\ \hline
MDiscTextClsEnt & 4.22             \\ \hline
MDiscTextEntCls & 0.54             \\ \hline
MFreqTextClsEnt & 7.71             \\ \hline
MFreqTextEntCLs & 0.72             \\ \hline
MDiscMathClsEnt & 4.16             \\ \hline
MDiscMathEntCls & 0.33             \\ \hline
MFreqMathClsEnt & 7.22             \\ \hline
MFreqMathEntCLs & 0.74             \\ \hline
\end{tabular}
\label{tab:textmathmdiscmfreqclsent}
\end{table}

Table \ref{tab:textmathmdiscmfreqclsent} shows the results of our examination. For the MDisc assessment, a LogReg classifier on TF-IDF encodings and LIME explainer is employed being chosen as best results. We interpret higher entropy as lower explainability because a low entropy leads to `sharp' class predictions. The results indicate that with text having a higher entropy than math, math features have a better explainability potential than text features. Moreover, with the entropy of MFreq being larger than for MDisc, the discriminative ranking seems to be more explainable than frequency ranking, underlining the power of the LIME model compared to a simple text frequency statistic. Finally, the observation that the entropy of classes-entities is higher than for entities-classes suggests that a prediction of classes from entities is more explainable than for classes from entities. This seems to be consistent with the fact that there are more entities than classes.

\section{Outlook}

In this final section, we recall our findings and outline future directions.

\subsection{Conclusion}

In this paper, we show that 1) the ambiguity of identifier symbols is distributed across document category classes while their semantic names are more predictable with smaller categorical cross-entropy; 2) an increasing number of unsupervised identifier name augmentation deteriorates the classification accuracy; 3) removing identifier names from the textual document features significantly reduces classification performance; 4) 95\% of the examined text entity N-grams could be linked to a Wikipedia article, and for 80\% of the examined formulae a concept name could be found within a surrounding text window of $\pm 10$ words; and 5) math features are potentially more explainable than text features and a discriminative ranking is more explainable than a frequency ranking in terms of entropy.

We answer our research question 1 by the entropy distribution for finding 1). Moreover, we answer research questions 2 and 3 by findings 2) and 3). Note that this means that identifier name augmentations seem to have some explanatory value, but MathEL for identifiers maybe not necessary for explainability as it deteriorates the classification accuracy. Finally, we answer research question 4 by findings 4) and research question 5 by findings 5). We conclude that while mathematical Entity Linking has some difficulties, i.e., some of the concept names are `hidden' in the context and maybe impossible to automatically detect, it can be a promising candidate for providing explainability in the case of `math-heavy' STEM documents.

\subsection{Future Work}

In the future, we will investigate how the definition of the math entity formula concept may affect MathEL results. More generally, we aim to further elaborate on a distinction between natural and mathematical language entities. We will try to cluster formula identifier category namespaces and assess cluster purity.

Moreover, we plan to develop Wikidata knowledge base encodings (`Bag-of-QIDs') for the textual and mathematical entities.
We will then seed the relations between arXiv and MSC categories to the Wikidata knowledge graph and persist entity-category linkings to evaluate the quality of knowledge graph labeling.
For fine-granular classification, there are often not enough training examples per class, as each is very small. Hence, a classification using knowledge-graph relations (linking entity keywords to subject classes) may be very beneficial to label small classes with little training data.  If expedient, we will also employ costly supervised entity and category annotation in an active learning framework, a process for which we first need to develop guidelines.

Besides for classification explainability and unsupervised labeling, Entity Linking can also be use for Semantic Search and Question Answering as well as document representation and ontology learning and summarization. Our long-term goal is to increase automation by unsupervised linking of textual and mathematical entities to serve these applications.



\bibliography{ref}{}
\bibliographystyle{unsrt}

\end{document}